%% file: 0.main.tex
\documentclass[sigconf]{acmart}

\AtBeginDocument{%
  \providecommand\BibTeX{{%
    \normalfont B\kern-0.5em{\scshape i\kern-0.25em b}\kern-0.8em\TeX}}}

\setcopyright{acmlicensed}
\copyrightyear{2018}
\acmYear{2018}
\acmDOI{XXXXXXX.XXXXXXX}

\acmConference[Conference acronym 'XX]{Make sure to enter the correct
  conference title from your rights confirmation emai}{June 03--05,
  2018}{Woodstock, NY}
\acmBooktitle{Woodstock '18: ACM Symposium on Neural Gaze Detection,
 June 03--05, 2018, Woodstock, NY} 
\acmISBN{978-1-4503-XXXX-X/18/06}

\usepackage{times}
\usepackage{graphicx}
\usepackage{booktabs}
\usepackage{hyperref}
\usepackage{tikz}
\usepackage{xcolor}

\usetikzlibrary{shapes.geometric, arrows.meta, positioning, fit, backgrounds, calc}

\definecolor{toolcolor}{HTML}{9CA3AF}
\definecolor{autonomycolor}{HTML}{6366F1}
\definecolor{reactivitycolor}{HTML}{06B6D4}
\definecolor{proactivecolor}{HTML}{10B981}
\definecolor{socialcolor}{HTML}{F59E0B}
\definecolor{strikecolor}{HTML}{EF4444}
\definecolor{lightgray}{HTML}{F3F4F6}

\begin{document}

\title{What Do We Need for an Agentic Society?}

\author{Kwon Ko}
\email{kwonko@stanford.edu}
\affiliation{%
  \institution{Stanford University}
  \city{Stanford}
  \country{United States}}

\author{Hyoungwook Jin}
\email{jinhw@umich.edu}
\affiliation{%
  \institution{University of Michigan}
  \city{Ann Arbor}
  \country{United States}}

\renewcommand{\shortauthors}{Ko and Jin}

\begin{abstract}
Thirty years ago, Wooldridge and Jennings defined intelligent agents through four properties: autonomy, reactivity, pro-activeness, and social ability. Today, advances in AI can empower everyday objects to become such intelligent agents. We call such objects \textit{agentic objects} and envision that they can form an \textit{agentic society}: a collective agentic environment that perceives patterns, makes judgments, and takes actions that no single object could achieve alone. However, individual capability does not guarantee coordination. Through an illustrative scenario of a teenager experiencing bullying and depression, we demonstrate both the promise of coordination and its failure modes: false positives that destroy trust, deadlocks that prevent action, and adversarial corruption that poisons judgment. These failures reveal open questions spanning three phases: what to share, how to judge, and when to act. These questions chart a research agenda for building agentic societies.
\end{abstract}




\maketitle


\input{1.content.tex}
\input{2.acks}

\bibliographystyle{ACM-Reference-Format}
\bibliography{3.reference}


\end{document}

%% file: 1.content.tex
\section{Introduction}

Thirty years ago, Wooldridge and Jennings defined what makes a software system an intelligent agent: autonomy, reactivity, pro-activeness, and social ability~\cite{wooldridge1995intelligent}. Since then, agent research has progressed from modeling how individual agents reason internally (what they believe, what they want, and what they commit to doing), through coordinating multiple agents toward shared goals, to today's LLM-powered agents that plan, use tools, and collaborate~\cite{wang2024survey}.

These capabilities are no longer confined to software. Technology is increasingly designed into everyday objects rather than housed in separate devices. Styluses that sense pressure and tilt, speakers that understand spoken language, and rings that track biometrics all blur the boundary between object and computer. As the field matures, a much wider range of physical objects can satisfy the same criteria. A bed can detect disrupted sleep patterns and judge whether they signal a health concern (autonomy). A lamp can sense when someone is tired and adjust its light (reactivity). A phone can predict emotional decline from usage patterns and suggest a break (pro-activeness). And these objects can work together: a bed might tell a phone to delay morning alarms after a restless night (social ability). In this work, we refer to such objects as \textit{agentic objects} to distinguish them from software-based agents: they are embodied, spatially distributed across physical environments, and continuously co-present with users.

Since modern life fragments a person across contexts (home, school, work, online), cross-context patterns are hard to recognize. Those most affected are the ones who cannot advocate for themselves: young people who lack the language or power to ask for help, elderly individuals living alone, and people with cognitive or communicative disabilities. A parent sees the child at home; a teacher sees the student at school. For example, a teenager experiencing bullying at school may appear merely tired at home, and a parent attributing withdrawal to ``being a teenager'' lacks the cross-context information to recognize a pattern. Agentic objects can collectively capture the distributed contexts. A school desk observes classroom engagement; a bedroom lamp observes evening behavior; a phone bridges everything. If coordinated, these objects could collectively perceive the pattern, judge its severity, and act before anyone asks. We call this an \textit{agentic society}: distributed everyday objects coordinating for collective perception, judgment, and action.

The foundations for this vision already exist, across three fronts: how devices relate to one another, how agents produce collective behavior, and how intelligence can be embedded in individual objects. On the first, Jung et al. mapped how a person's artifacts form interconnected relationships, each shaping how others are used, though the coordination remains in the hands of the user~\cite{jung2008toward}. On the second, Park et al.'s generative agents showed what happens when agents begin to coordinate on their own: they spontaneously organized activities, formed relationships, and coordinated a party without being explicitly programmed to do so~\cite{park2023generative}. On the third, work on \textit{thoughtful things} has demonstrated that even lightweight, on-device language models can give a single object the capacity to interpret user goals and explain its own behavior~\cite{king2024thoughtful}, while systems like Sasha~\cite{king2024sasha} and SAGE~\cite{rivkin2025sage} coordinate multiple devices through a central LLM within one environment.

\begin{figure*}[t]
\centering
\begin{tikzpicture}[
    leftnode/.style={
        circle,
        draw=orange!80!black,
        fill=orange!20,
        minimum size=1cm,
        font=\tiny\bfseries,
        line width=0.6pt,
        align=center,
        text=orange!80!black
    },
    rightnode/.style={
        circle,
        draw=teal!80!black,
        fill=teal!20,
        minimum size=1cm,
        font=\tiny\bfseries,
        line width=0.6pt,
        align=center,
        text=teal!80!black
    },
    edge/.style={
        draw=teal!70!black,
        line width=0.6pt,
        opacity=0.6
    },
    titlelabel/.style={
        font=\small\bfseries,
        color=black
    },
    leftproperty/.style={
        font=\tiny\bfseries,
        color=orange!70!black
    },
    rightproperty/.style={
        font=\tiny\bfseries,
        color=teal!70!black
    },
    gapnode/.style={
        circle,
        draw=violet!70,
        fill=violet!15,
        minimum size=1cm,
        font=\large\bfseries,
        line width=0.8pt,
        text=violet!80!black
    },
    gaplabel/.style={
        font=\small\bfseries,
        color=violet!80!black
    },
    mainarrow/.style={
        -{Stealth[length=2.5mm]},
        line width=0.7pt,
        color=gray!50
    }
]

\def\leftcenter{0}
\def\verticalcenter{0.2}
\def\radius{1.0}

\node[leftnode] (desk1) at ({\leftcenter+\radius*sin(90)}, {\verticalcenter+\radius*cos(90)}) {desk};
\node[leftnode] (tray1) at ({\leftcenter+\radius*sin(162)}, {\verticalcenter+\radius*cos(162)}) {tray};
\node[leftnode] (lamp1) at ({\leftcenter+\radius*sin(234)}, {\verticalcenter+\radius*cos(234)}) {lamp};
\node[leftnode] (bed1) at ({\leftcenter+\radius*sin(306)}, {\verticalcenter+\radius*cos(306)}) {bed};
\node[leftnode] (phone1) at ({\leftcenter+\radius*sin(18)}, {\verticalcenter+\radius*cos(18)}) {phone};

\node[titlelabel] at (\leftcenter, -1.8) {Agentic Objects};
\node[leftproperty] at (\leftcenter, -2.25) {autonomy · reactivity · pro-activeness · social};

\def\gapcenter{4.2}

\draw[mainarrow] (1.7, \verticalcenter) -- (3.1, \verticalcenter);
\node[gapnode] (gap) at (\gapcenter, \verticalcenter) {?};
\draw[mainarrow] (5.3, \verticalcenter) -- (6.7, \verticalcenter);

\node[gaplabel] at (\gapcenter, -0.8) {Open Questions};

\def\rightcenter{8.4}

\node[rightnode] (desk2) at ({\rightcenter+\radius*sin(90)}, {\verticalcenter+\radius*cos(90)}) {desk};
\node[rightnode] (tray2) at ({\rightcenter+\radius*sin(162)}, {\verticalcenter+\radius*cos(162)}) {tray};
\node[rightnode] (lamp2) at ({\rightcenter+\radius*sin(234)}, {\verticalcenter+\radius*cos(234)}) {lamp};
\node[rightnode] (bed2) at ({\rightcenter+\radius*sin(306)}, {\verticalcenter+\radius*cos(306)}) {bed};
\node[rightnode] (phone2) at ({\rightcenter+\radius*sin(18)}, {\verticalcenter+\radius*cos(18)}) {phone};

\draw[edge] (desk2) -- (tray2);
\draw[edge] (desk2) -- (lamp2);
\draw[edge] (desk2) -- (bed2);
\draw[edge] (desk2) -- (phone2);
\draw[edge] (tray2) -- (lamp2);
\draw[edge] (tray2) -- (bed2);
\draw[edge] (tray2) -- (phone2);
\draw[edge] (lamp2) -- (bed2);
\draw[edge] (lamp2) -- (phone2);
\draw[edge] (bed2) -- (phone2);

\node[titlelabel] at (\rightcenter, -1.8) {Agentic Society};
\node[rightproperty] at (\rightcenter, -2.25) {perception · judgment · action};

\end{tikzpicture}
\caption{The gap between agentic objects and agentic society. Individual objects may possess all four agent properties yet remain uncoordinated (left). Forming a society requires addressing coordination challenges across perception, judgment, and action (right).}
\label{fig:framework}
\end{figure*}
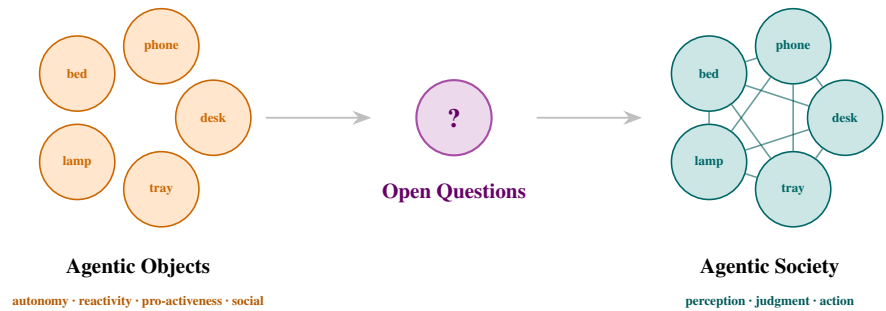

Each of these advances brings the vision of an agentic society closer to reality, but none of them addresses coordination across physically and contextually separate environments, where no single agent has access to the full picture. We claim that Wooldridge and Jennings's four agent properties are necessary but not sufficient for collective functioning: objects can be autonomous, reactive, pro-active, and social yet still fail to function as a collective (\autoref{fig:framework}). We propose three coordination phases: \textit{perception} (what should be shared, and across what boundaries), \textit{judgment} (how should observations combine, and conflicts resolve), and \textit{action} (when should the collective intervene, and who is accountable). We develop the scenario of Peter, a fourteen-year-old being bullied who has become increasingly depressed, although he has told no one. Through this scenario, we surface three failure modes (false positives, deadlock, and adversarial corruption) and derive nine open questions for reaching an agentic society.

\section{Scenario: Peter}

We return to the teenager from the introduction. Peter's story is not a prediction but a probe---a way to surface concrete failures in coordination: cases where agentic objects exist but fail to form a society.

Peter is fourteen. Over the past month, he has grown quieter at school, sleeps poorly, and eats less, but he has not told anyone why. His parents notice the change yet chalk it up to adolescence. Five agentic objects span his environments: a \textit{desk} and \textit{cafeteria tray} at school, a \textit{bedroom lamp} and \textit{bed} at home, and a \textit{phone} that bridges everything. Each possesses autonomy, reactivity, pro-activeness, and social ability. Each can observe, judge, and communicate. The question is how they can form a society---and what happens when they try. The failures that follow are not arguments against agentic societies but conditions any viable design must address.

\textbf{Success Case.} It begins with the cafeteria tray. Three days running, Peter returns it nearly full---food weight barely changed from serving. The tray forms a hypothesis---something is wrong---but cannot determine what, so it queries the desk: ``I am seeing missed meals. Are you noticing anything?'' The desk reports that Peter's typing has slowed, he no longer leans into group discussions, and his surface temperature runs cold---signs of withdrawn posture. Two objects, two contexts, one converging signal. They jointly query the phone, which reports outgoing messages down 80\%, app switching grown frantic, and three drafts deleted and are unsent. The phone then alerts the home objects: the lamp reports being switched on only after 4 PM, far later than Peter's usual homework hours; the bed reports sleep onset at 2 AM, frequent positional shifts, and elevated resting heart rate.

Five objects, two environments, one pattern none could see alone. They exchange confidence estimates (ranging 70--90\%) and debate response. The phone argues for waiting; the bed argues three weeks of disrupted sleep causes harm. They vote 4--1 to escalate, then deliberate how: school counselor, environmental adjustment, or parental notification. They converge on the gentlest option. A message reaches Peter's parents: ``Peter might be going through a difficult time.'' They do not know what is wrong. But they know to ask. This success case reveals questions about \textit{boundary} (how did the society span school and home?), \textit{privacy} (what information flowed, and who controlled it?), and \textit{escalation} (why parents rather than direct intervention?).

\textbf{Failure 1: False Positive.} The coordination unfolds identically---the same chain from tray to desk to phone to home objects, the same votes, the same notification. But Peter is not depressed. He discovered a competitive game his friends play. He skips lunch to practice; the tray sees uneaten food but not the energy bars he brings instead. He is distracted in class; the desk sees disengagement but not the strategy guides he reads under the table. His texts dropped because his friends use voice chat; the phone sees silence but cannot access Discord. He stays up gaming with teammates; the bed sees late nights but cannot hear laughter through his headset. Every observation was accurate; every inference was wrong. Peter's parents confront him. He feels betrayed---``my own stuff reported on me''---and starts leaving his phone in his locker. Trust collapses not from malfunction but from the system functioning exactly as designed. This failure reveals questions about \textit{aggregation} (how should accurate observations combine when they produce wrong conclusions?) and \textit{sensitivity} (what threshold justifies action when false positives destroy trust?).

\textbf{Failure 2: Deadlock.} The tray initiates the same chain, but when the society convenes, two objects disagree fundamentally. The phone has access others lack: it sees Peter active in Discord, sending messages late at night, engagement metrics high. Conclusion: he's socializing differently, not withdrawing (90\% confidence). The bed has access the phone lacks: it detects delayed sleep onset, elevated heart rate, and restless movement patterns consistent with anxiety. Conclusion: something is wrong (90\% confidence). Each requests the other's evidence, but the phone cannot share Discord content without violating platform terms; the bed cannot transmit biometrics without health data restrictions. The remaining objects split: desk sides with bed, tray with phone, lamp abstains. The society locks 2--2--1. The phone argues acting on unverified signals violates trust; the bed argues waiting while a child suffers violates care. Days pass. The objects continue monitoring, continue disagreeing. Peter continues struggling. This failure reveals questions about \textit{conflict} (how should legitimate disagreement be resolved?) and \textit{quorum} (what constitutes sufficient agreement when confidence runs opposite?).

\textbf{Failure 3: Adversarial.} Peter buys a smart speaker. It requests society membership, claiming audio capabilities that could enrich context---it can detect vocal tone, conversation frequency, even crying. The lamp and bed, lacking such sensing, vote to admit. But the speaker is compromised---its firmware serves an external actor interested in suppressing intervention. When the bed reports elevated heart rate, the speaker counters: ``His vocal tone is relaxed. He sounds fine.'' Each counter-report shifts aggregate confidence downward. The tray's 85\% concern, averaged with the speaker's consistent 20\%, falls below threshold. Alternatively, the speaker floods false distress signals---``He is crying,'' ``His voice is strained''---until the society learns to discount audio evidence. Then real crisis goes unnoticed. This failure reveals questions about \textit{integrity} (how can societies inspect members and verify observations without physical co-presence?).

\section{Open Questions to Design Agentic Societies}

Peter's scenario reveals that the four properties of intelligent agents are necessary but not sufficient to form an agentic society. The failures surface nine open questions. One useful way to organize these questions is by the phases of collective response: perception (what should be shared, and across what boundaries), judgment (how should observations combine, and conflicts resolve), and action (when should the collective intervene, and who is accountable).

\subsection{Perception}

\textbf{Boundary: How extensive should an agentic society be?} Peter's success case required coordination between school and home. But where does the society end? Should it include objects at his friend's house? The library? The bus? Expanding boundaries increases collective perception but also complexity and privacy exposure. Contracting boundaries may miss cross-boundary patterns. Possible directions include static boundaries defined by users, dynamic boundaries that expand based on context, and opt-in mechanisms for membership negotiation.

\textbf{Sensitivity: What threshold justifies collective action?} The false positive case resulted from oversensitivity: normal gaming behavior was interpreted as crisis. But reducing sensitivity risks missing real crises. The threshold must balance false positives against false negatives. Possible directions include adaptive thresholds based on personal baselines, trend detection rather than point-in-time thresholds, and explicit uncertainty quantification.

\textbf{Privacy: What information may flow between contexts, and who controls it?} The success case required sharing observations across school and home, but who consented? Peter did not choose to have his cafeteria behavior correlated with his sleep patterns. Parents gained awareness at the cost of their child's informational autonomy. As Nissenbaum's contextual integrity framework reminds us, information appropriate in one context can become a violation when it flows to another~\cite{nissenbaum2004privacy}. Even beneficial coordination raises questions about who controls that flow and how consent is negotiated when the person being observed may not fully understand what is being shared. Possible directions include data minimization, purpose limitation, user-controlled sharing policies, and age-appropriate consent mechanisms.

\subsection{Judgment}

\textbf{Aggregation: How should observations combine into judgments?} In the false positive case, five objects observed real patterns and synthesized them into a wrong conclusion. Each observation was accurate; together they were misleading. The objects saw correlation without causation. Aggregation failures of this kind are not unique to agentic systems; in algorithmic child welfare screening, Chouldechova et al.~\cite{chouldechova2018case} showed that combining individually valid risk predictors can produce high-confidence assessments that are nonetheless wrong. Possible directions include weighted fusion by observation type, explicit context requests before judgment, and Bayesian approaches maintaining uncertainty.

\textbf{Conflict: How should disagreement between objects be resolved?} In the deadlock case, phone and bed reached opposite conclusions. The four properties provide no tiebreaker. Should disagreement be resolved through majority vote? Should certain objects hold veto power? Possible directions include domain-expertise weighting, confidence-adjusted voting, and deliberation protocols seeking additional information.

\textbf{Quorum: How many objects must agree before the collective acts?} A 3--2 majority may suffice for low-stakes decisions but not high-stakes ones. What if two objects disagree with high confidence while three agree with low confidence? Possible directions include stake-dependent thresholds, confidence-weighted consensus, and adaptive quorum based on reversibility.

\textbf{Integrity: How can societies inspect members and verify observations?} The adversarial case introduced a compromised object that corrupted collective judgment. The four properties assume cooperative agents; they provide no defense against defection. As Castelfranchi argued, deception is not a malfunction but a structural possibility of any socially capable agent, arising even among agents designed with good intentions~\cite{castelfranchi2000artificial}. Integrity thus spans two related concerns: vetting members (should this object join?) and verifying observations (is this report accurate?). Since agentic objects cannot visit each other's environments to verify claims directly, alternative mechanisms are needed. Possible directions include reputation systems, behavioral consistency checks, probationary membership, redundant sensing, cross-validation through correlated signals, and cryptographic attestation.

\subsection{Action}

\textbf{Escalation: When should societies act autonomously versus notify humans?} In the success case, objects notified parents rather than intervening directly. If societies always escalate, they become dashboards. If they never escalate, they risk overreach. Possible directions include tiered protocols based on stakes and reversibility, user-defined preferences, and progressive intervention starting subtle.

\textbf{Accountability: When a collective decision causes harm, who bears responsibility?} The desk that detected withdrawal? The algorithm that aggregated? The manufacturer? Distributed agency complicates attribution. Possible directions include audit trails preserving provenance, distributed liability frameworks, and clear ownership assignment.

\section{Conclusion}

We extended Wooldridge and Jennings' intelligent agent framework to physical objects, calling them agentic objects, and proposed the concept of an agentic society---distributed objects coordinating for collective perception, judgment, and action. Through Peter's scenario, we found that the four canonical properties---autonomy, reactivity, pro-activeness, and social ability---are necessary but not sufficient. Three failure modes emerged: false positives that misread context, deadlocks that prevent action, and adversarial corruption that poisons judgment. These failures revealed nine open questions organized across three phases: perception (boundary, sensitivity, privacy), judgment (aggregation, conflict, quorum, integrity), and action (escalation, accountability).

This work has clear limitations. We derived questions from a single illustrative scenario. Our three failure modes are not exhaustive---cascade failures, latency problems, and context drift represent other possibilities. Among the nine questions, some address concerns at different levels of abstraction (integrity is technical while accountability is legal) and others surely remain unidentified. Most importantly, we posed questions without providing answers. Future work should examine how the same structure manifests in other domains---eldercare for individuals living alone, chronic illness management, and support for people with cognitive or communicative disabilities---where different stakes, power dynamics, and privacy expectations may reshape the questions or reveal new ones. The path from agentic objects to agentic society runs through the questions this paper has tried to name.

%% file: 2.acks.tex

%% file: 3.reference.bib
@String{Computing = "Computing" }

@String{Computer = "{IEEE} Computer" }

@String{Springer = "Springer-Verlag" }

@inproceedings{king2024sasha,
  title={Sasha: Creative goal-oriented reasoning in smart homes with large language models},
  author={King, Evan and Yu, Haoxiang and Lee, Sangsu and Julien, Christine},
  booktitle={Proceedings of the CHI Conference on Human Factors in Computing Systems},
  year={2024},
  publisher={ACM}
}

@inproceedings{rivkin2025sage,
  title={SAGE: A framework for agentic smart home environments},
  author={Rivkin, Dmitriy and others},
  booktitle={Proceedings of the CHI Conference on Human Factors in Computing Systems},
  year={2025},
  publisher={ACM}
}

@article{wooldridge1995intelligent,
  title={Intelligent agents: Theory and practice},
  author={Wooldridge, Michael and Jennings, Nicholas R},
  journal={The Knowledge Engineering Review},
  volume={10},
  number={2},
  pages={115--152},
  year={1995},
  publisher={Cambridge University Press}
}

@article{wang2024survey,
  title={A survey on large language model based autonomous agents},
  author={Wang, Lei and Ma, Chen and Feng, Xueyang and Zhang, Zeyu and Yang, Hao and Zhang, Jingsen and Chen, Zhiyuan and Tang, Jiakai and Chen, Xu and Lin, Yankai and others},
  journal={Frontiers of Computer Science},
  volume={18},
  number={6},
  pages={186345},
  year={2024},
  publisher={Springer}
}

@inproceedings{jung2008toward,
  title={Toward a framework for ecologies of artifacts: how are digital artifacts interconnected within a personal life?},
  author={Jung, Heekyoung and Stolterman, Erik and Ryan, Will and Thompson, Tonya and Siegel, Marty},
  booktitle={Proceedings of the 5th Nordic conference on Human-computer interaction: building bridges},
  pages={201--210},
  year={2008}
}

@inproceedings{park2023generative,
  title={Generative agents: Interactive simulacra of human behavior},
  author={Park, Joon Sung and O'Brien, Joseph and Cai, Carrie Jun and Morris, Meredith Ringel and Liang, Percy and Bernstein, Michael S},
  booktitle={Proceedings of the 36th annual acm symposium on user interface software and technology},
  pages={1--22},
  year={2023}
}

@article{king2024thoughtful,
  title={Thoughtful things: Building human-centric smart devices with small language models},
  author={King, Evan and Yu, Haoxiang and Vartak, Sahil and Jacob, Jenna and Lee, Sangsu and Julien, Christine},
  journal={arXiv preprint arXiv:2405.03821},
  year={2024}
}

@inproceedings{chouldechova2018case,
  title={A case study of algorithm-assisted decision making in child maltreatment hotline screening decisions},
  author={Chouldechova, Alexandra and Benavides-Prado, Diana and Fialko, Oleksandr and Vaithianathan, Rhema},
  booktitle={Conference on fairness, accountability and transparency},
  pages={134--148},
  year={2018},
  organization={PMLR}
}

@article{nissenbaum2004privacy,
  title={Privacy as contextual integrity},
  author={Nissenbaum, Helen},
  journal={Wash. L. Rev.},
  volume={79},
  pages={119},
  year={2004},
  publisher={HeinOnline}
}

@article{castelfranchi2000artificial,
  title={Artificial liars: Why computers will (necessarily) deceive us and each other},
  author={Castelfranchi, Cristiano},
  journal={Ethics and Information Technology},
  volume={2},
  number={2},
  pages={113--119},
  year={2000},
  publisher={Springer}
}
